\documentclass{aa}
\usepackage{graphicx}
\usepackage{natbib}

\bibpunct{(}{)}{;}{a}{}{,}

\newcommand{\GILDAS}{\texttt{GILDAS}}

\newcommand{\IRAM}{\textrm{IRAM}}
\newcommand{\IRAMthm}{\textrm{IRAM-30m}}
\newcommand{\PdBI}{\textrm{PdBI}}
\newcommand{\CSO}{\textrm{\CSO}}

\newcommand{\ie} {{\em i.e.}}
\newcommand{\eg} {{\em e.g.}}

\newcommand{\HH}    {\mbox{H$_2$}}           
\newcommand{\DD}      {\mbox{D$_2$}}         
\newcommand{\HHHp}    {\mbox{H$_3^+$}}       
\newcommand{\HHDp}    {\mbox{H$_2$D$^+$}}    
\newcommand{\HDDp}    {\mbox{HD$_2^+$}}      

\newcommand{\CHHHp}   {\mbox{CH$_3^+$}}      

\newcommand{\CeiO}  {\mbox{C$^{18}$O}}       
\newcommand{\DCOp}  {\mbox{DCO$^{+}$}}       
\newcommand{\HCOp}  {\mbox{HCO$^{+}$}}       
\newcommand{\HthCOp}{\mbox{H$^{13}$CO$^{+}$}}
\newcommand{\DDCO}  {\mbox{D$_2$CO}}         
\newcommand{\NHDD}  {\mbox{NHD$_2$}}         

\newcommand{\Jone}{\mbox{$J$=1--0}}
\newcommand{\Jtwo}{\mbox{$J$=2--1}}
\newcommand{\Jthr}{\mbox{$J$=3--2}}
\newcommand{\Jfiv}{\mbox{$J$=5--4}}

\newcommand{\emm}[1]{\ensuremath{#1}}   
\newcommand{\emr}[1]{\emm{\mathrm{#1}}} 
\newcommand{\unit}[1]{\emm{\, \emr{#1}}}
\newcommand{\K}   {\unit{K}}
\newcommand{\mm}  {\unit{mm}}

\newcommand{\pscm}{\unit{cm^{-2}}}
\newcommand{\pccm}{\unit{cm^{-3}}}
\newcommand{\Kpccm}{\unit{K\,cm^{-3}}}

\newcommand{\kms}   {\unit{km\,s^{-1}}}

\newcommand{\pc}    {\unit{pc}}

\newcommand{\Av}{\emm{A_V}}

\renewcommand{\deg}{\emm{^\circ}}

\newcommand{\nds}[1]{\emm{\displaystyle#1}} 
\newcommand{\cbrace}[1] {\nds{\left\{ #1 \right\}}} 

\newcommand{\Tas}{\emm{T_\emr{A}^*}}
\newcommand{\Tmb}{\emm{T_\emr{mb}}}
\newcommand{\Beff}{\emm{B_\emr{eff}}}
\newcommand{\Feff}{\emm{F_\emr{eff}}}

\newcommand{\about}{\emm{\sim}}

\newcommand{\TabObs}{%
  \begin{table*}
    \caption{Observation parameters. The projection center of all the data is $\alpha_{2000} = 05^h40^m54.27^s$, $\delta_{2000} = -02\deg 28' 00''$.}
    {\tiny
    \begin{tabular}{rcrlcccrclcr}
      \hline \hline
      Molecule & Transition & Frequency  & Instrument & \# Pix.$^{a}$ & \Feff{}$^{a}$ & \Beff{}$^{a}$ & Resol. & Resol. & Int. Time$^{a,b}$ & Noise$^{c}$ & Obs. date$^{a}$ \\
      &            & GHz        &            &         &         &         & arcsec & \kms{} & hours     &      K      &     \\
      \hline
      \HthCOp{} & \Jthr{} & 260.255339 & 30m/HERA  & 9 & 0.90 & 0.46 & $13.5''$ & 0.20 & 5.9/11.3 & 0.06 & Mar. 2006 \\
      \HthCOp{} & \Jone{} &  86.754288 & 30m+PdBI  & 2 & 0.95 & 0.78 & $ 6.7''$ & 0.20 & 2.6/4.5  & 0.10 & Sep. 2006 \\
        \DCOp{} & \Jthr{} & 216.112582 & 30m/HERA  & 9 & 0.90 & 0.52 & $11.4''$ & 0.11 & 1.5/2.0  & 0.10 & Mar. 2006 \\
        \DCOp{} & \Jtwo{} & 144.077289 & 30m/CD150 & 2 & 0.93 & 0.69 & $18.0''$ & 0.08 & 5.9/8.7  & 0.18 & Sep. 2006 \\
      \hline
      \CeiO{} & \Jtwo{} & 219.560319          & 30m/HERA  & 9   & 0.91 & 0.55 & $11.2''$ & 0.11 & -- & 0.26 & May 2003 \\
      \multicolumn{3}{c}{Continuum at 1.2\mm} & 30m/MAMBO & 117 &   -- & --   & $11.7''$ &   -- & -- &   -- &       -- \\
      \hline
    \end{tabular}}\\
    $^{a}$ Those columns apply to the 30m data but not to the PdBI data for
    the \HthCOp{}~\Jone{} line. $^{b}$ Two values are given for the
    integration time: the on-source time and the telescope time. $^{c}$
    Noise values estimated at the position of the \DCOp{} peak. 
    \label{tab:obs}
  \end{table*}}

\newcommand{\TabTrans}{%
  \begin{table}
    \caption{Einstein coefficients, upper level energies and critical densities 
      for the range of temperatures considered in this work.}
    \begin{center}
      \begin{tabular}{lccrc} 
        \hline \hline
        Molecule  & Transition &       $A_{ij}$          &  $E_\emr{up}$  &    $n_\emr{crit}$     \\
                  &            &      (s$^{-1}$)         &      (K)       &
        (cm$^{-3}$)       \\
        \hline
        \HthCOp{} & \Jone{}    &  $3.9 \times 10^{-5}$   &     ~4.2       & $\sim2 \times 10^{5}$ \\
        \HthCOp{} & \Jthr{}    &  $1.3 \times 10^{-3}$   &     25.0       & $\sim3 \times 10^{6}$ \\
        \DCOp{}   & \Jtwo{}    &  $2.1 \times 10^{-4}$   &     10.4       & $\sim6 \times 10^{5}$ \\
        \DCOp{}   & \Jthr{}    &  $7.7 \times 10^{-4}$   &     20.7       & $\sim2 \times 10^{6}$ \\ \hline
      \end{tabular}
      \label{tab:n_cr}
    \end{center} 
  \end{table}}

\newcommand{\FigMaps}{%
\begin{figure}
  \centering %
  \includegraphics[width=\hsize{}]{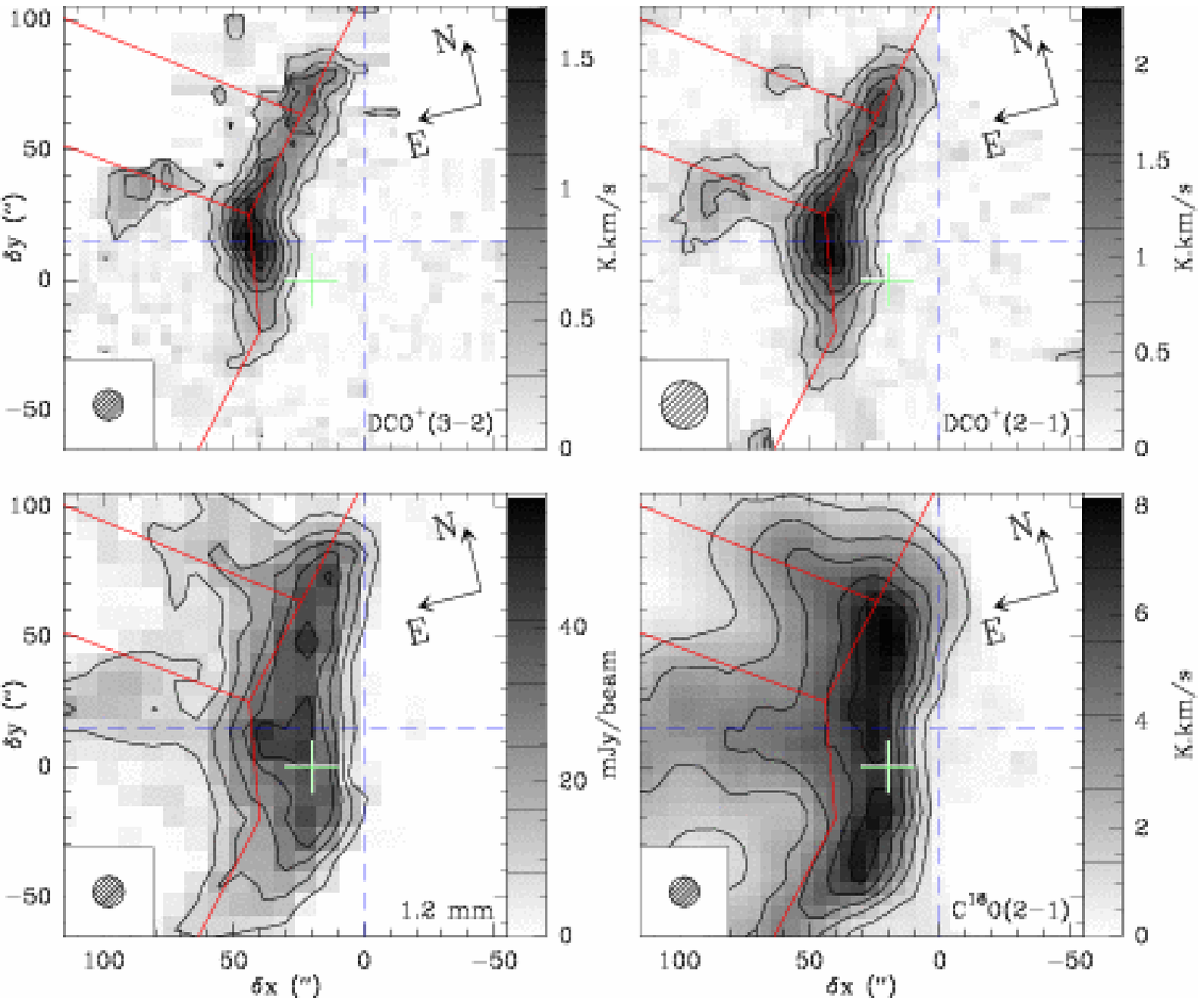} %
   \caption{\IRAMthm{} integrated intensity maps. Maps have been
     rotated by 14\deg{} counter--clockwise around the projection center,
     shown as the green cross at $(\delta x,\delta y)$ = $(20'',0'')$, to
     bring the exciting star direction in the horizontal direction and the
     horizontal zero has been set at the PDR edge, delineated by the dashed
     blue vertical line. The synthesized beam is plotted in the bottom left
     corner. Values of contour levels are shown on each image lookup table.
     The emission of all lines is integrated between 10.1 and 11.1\kms{}.}
  \label{fig:maps}
\end{figure}}

\newcommand{\FigCuts}{%
\begin{figure}
  \centering %
  \includegraphics[width=\hsize{},angle=270]{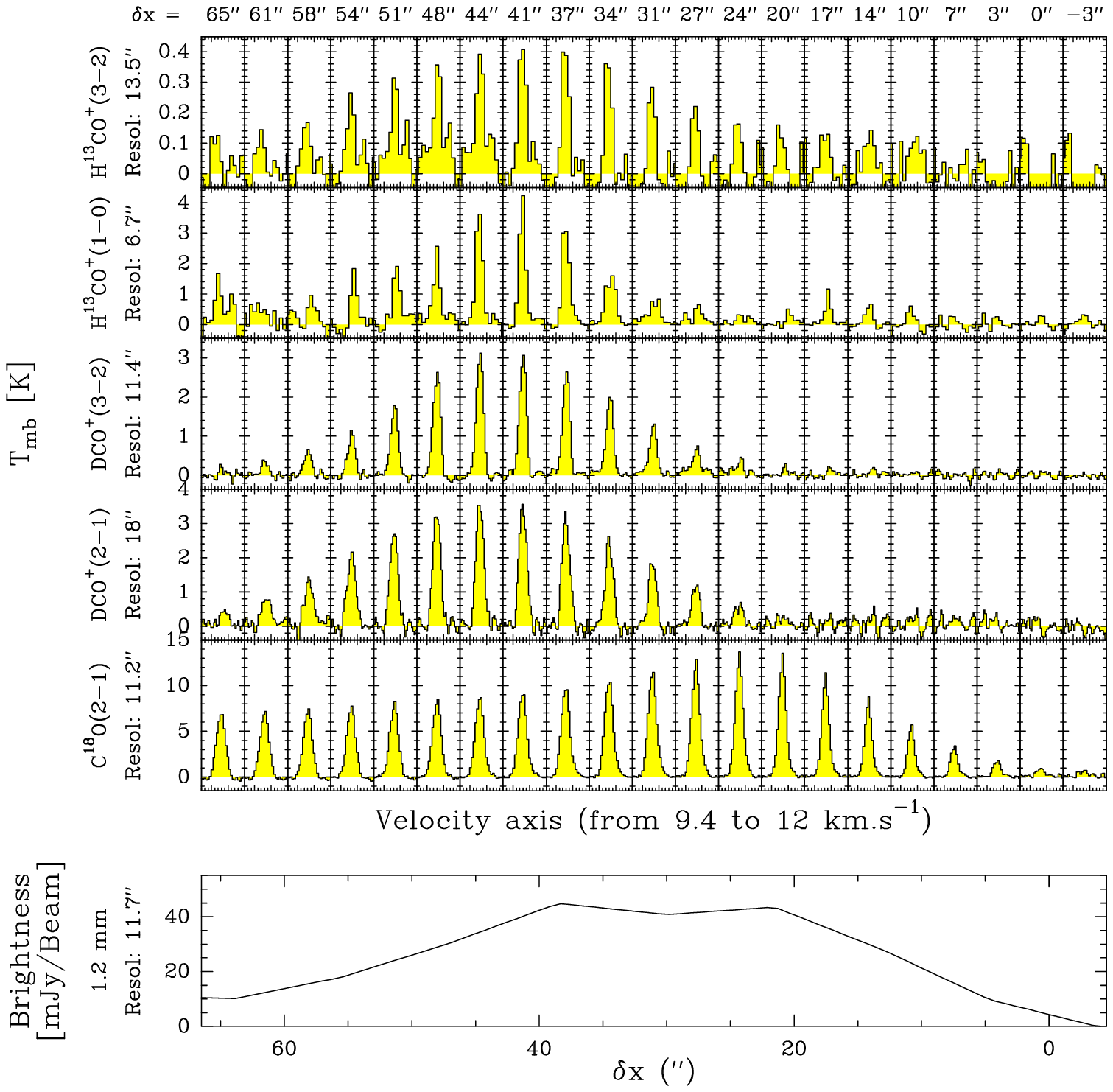} %
  \caption{Cut along the direction of the exciting star at 
    $\delta y = 15''$.}
  \label{fig:cuts}
\end{figure}}

\newcommand{\FigChem}{%
\begin{figure}
  \centering %
  \includegraphics[width=6.5cm,angle=270]{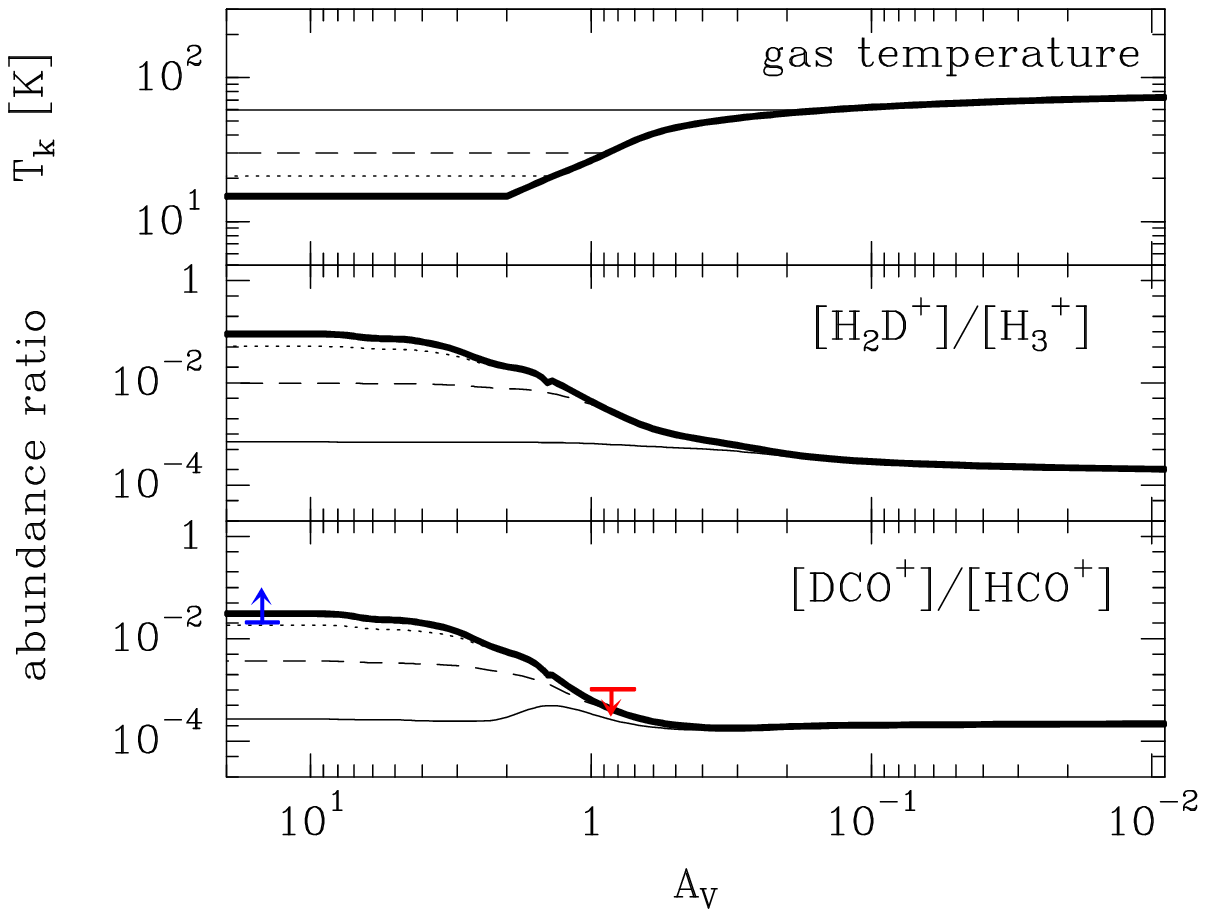} %
  \caption{Chemical models for different minimum
    gas temperatures: 15, 20, 30 and 60\K{}.  The density is
    $n_\emr{H}$=4$\times$10$^5$\pccm and the illuminating radiation field
    is $\chi=60$. Temperature profiles and predicted [\HHDp{}]/[\HHHp{}]
    and [\DCOp{}]/[\HCOp{}] abundance ratios are shown as a function of
    \Av{}. The [\DCOp{}]/[\HCOp{}] ratios inferred from observations in the
    cold condensation at $\delta x \sim 40-45''$ and in the warm PDR gas at
    $\delta x \sim 10-15''$ are shown respectively with the blue and red
    arrows.}
  \label{fig:chem}
\end{figure}}

\begin{document}

\title{Deuterium fractionation in the Horsehead edge\thanks{Based on
    observations obtained with the IRAM Plateau de Bure interferometer and
    30~m telescope. IRAM is supported by INSU/CNRS (France), MPG (Germany),
    and IGN (Spain).}}

\author{J. Pety \inst{1,2}%
\and J.R. Goicoechea \inst{2}%
\and P. Hily-Blant \inst{1}%
\and M. Gerin \inst{2}%
\and D. Teyssier \inst{3}%
}

\offprints{\email{pety@iram.fr}}

\institute{%
  IRAM, 300 rue de la Piscine, 38406 Grenoble cedex, France. \email{pety@iram.fr,hilyblan@iram.fr} %
  \and LERMA, UMR 8112, CNRS, Observatoire de Paris and Ecole Normale
  Sup\'erieure, 24 Rue Lhomond, 75231 Paris cedex 05, France.
  \email{javier@lra.ens.fr, gerin@lra.ens.fr} %
  \and European Space Astronomy Centre, Urb. Villafranca del Castillo, P.O.
  Box 50727, Madrid 28080, Spain.\\
  \email{dteyssier@sciops.esa.int} %
}

\date{Received 22 December 2006 / accepted 22 January 2007}

\abstract
{Deuterium fractionation is known to enhance the [\DCOp{}]/[\HCOp{}]
  abundance ratio over the D/H$\sim$10$^{-5}$ elemental ratio in the cold
  and dense gas typically found in pre-stellar cores.}
{We report the first detection and mapping of very bright \DCOp{}~\Jthr{}
  and~\Jtwo{} lines (3 and 4\K{} respectively) towards the Horsehead
  photodissociation region (PDR) observed with the \IRAMthm{} telescope.
  The \DCOp{} emission peaks close to the illuminated warm edge of the
  nebula ($<50''$ or $\about0.1\pc$ away).}
{Detailed nonlocal, non--LTE excitation and radiative transfer analyses
  have been used to determine the prevailing physical conditions and to
  estimate the \DCOp{} and \HthCOp{} abundances from their line
  intensities.}
{A large [\DCOp]/[\HCOp] abundance ratio $(\geq 0.02)$ is inferred at the
  \DCOp{} emission peak, a condensation shielded from the illuminating
  far-UV radiation field where the gas must be cold (10--20\K{}) and dense
  ($\geq 2 \times 10^5\pccm$). \DCOp{} is not detected in the warmer
  photodissociation front, implying a lower [\DCOp]/[\HCOp] ratio ($<
  10^{-3}$).}
{According to our gas phase chemical predictions, such a high deuterium
  fractionation of \HCOp{} can only be explained if the gas temperature is
  below 20\K{}, in good agreement with \DCOp{} excitation calculations.}

\keywords{{ISM clouds -- molecules -- individual object (Horsehead nebula)
    -- radio lines: ISM}}

\maketitle{}

Molecules are enriched in deuterium over the elemental D/H abundance
\citep[$1-2 \times 10^{-5}$,][]{linsky06} in many different astrophysical
environments. These include cold, dense cores~\citep{guelin82}, mid--planes
of circumstellar disks~\citep{vD03,guilloteau06}, hot molecular
cores~\citep{hatchell98}, and even PDRs~\citep{leurini06}. Multiply
deuterated species were first detected several years ago, \eg{} \DDCO{} in
warm gas~\citep{turner90} and \NHDD{} in cold gas~\citep{roueff00}.
\citet{solomon73} and~\citet{watson74} first proposed that deuterium
fractionation is mostly caused by gas-phase ion--molecule reactions.
\citet{smith82} and~\citet{roberts00a} confirmed that the deuteration of
\HHHp{} at low temperatures ($< 25\K$) and of \CHHHp{} at higher
temperatures (up to $\about 70\K$) are important precursor reactions in the
subsequent deuteration of other species.  \citet{roberts00b},
\citet{walmsley04} and \citet{flower06} succeeded in reproducing the amount
of several multiply deuterated molecules in cold gas by adding to pure
gas-phase chemistry the accretion (freeze-out) of gas-phase molecules onto
the surfaces of dust grains.  Finally the observed deuterium fractionation
in hot cores is thought to result from the liberation of deuterated
molecules, trapped in ice mantles in the prestellar phase.

Although it has been studied thoroughly for 30 years, deuterium chemistry
is not yet fully understood. With many chemical and physical processes
competing for efficient fractionation, models are easier to compare with
observations for sources with well described physical conditions.  In this
letter, we report the detection of very bright \DCOp{} lines in the
Horsehead edge.  In particular, the mane of the Horsehead nebula is a PDR
viewed nearly edge-on (inclination $< 5\deg$) illuminated by the O9.5V star
$\sigma$Ori~\citep{abergel03,philipp06}. \citet{habart05} showed that the
PDR has a very steep gradient, rising to $n_\emr{H} \sim 2 \times
10^5\pccm$ in less than $10''$ or 0.02\pc{}, at a roughly constant thermal
pressure of $\sim 4\times 10^6\Kpccm$.  The newly detected \DCOp{} lines
arise from a condensation adjacent to the PDR, first detected
by~\citet{hilyblant05}.  According to its submillimeter continuum emission,
this core (B33-SMM1) is $0.13 \times 0.31\pc$ long and has an average \HH{}
density of $\sim10^4\pccm$ and a peak density of $\sim 6\times 10^5\pccm$
\citep{wardthompson06}.

\FigMaps{} %
\TabObs{} %

\section{Observations and data reduction}
\label{sec:obs}

The \DCOp{}~\Jthr{} line was observed during 2 hours of excellent winter
weather ($\about 0.7\mm$ of water vapor) using the first polarization
(\ie{} nine of the eighteen available pixels) of the \IRAMthm{}/HERA single
sideband multi-beam receiver.  We used the frequency-switched, on-the-fly
observing mode. We observed along and perpendicular to the direction of the
exciting star in zigzags (\ie{} $\pm$ the lambda and beta scanning
direction). The multi-beam system was rotated by 9.6\deg{} with respect to
the scanning direction.  This ensured Nyquist sampling between the rows
except at the edges of the map.  The \DCOp{}~\Jtwo{} was observed during
11.3 hours using the C150 and D150 single-side band receivers of the
\IRAMthm{} under $\about 8.5\mm$ of water vapor.  We used the
frequency-switched, on-the-fly observing mode over a $160''\times 170''$
portion of the sky. Scanned lines and rows were separated by $8''$ ensuring
Nyquist sampling. A detailed description of the \CeiO{}~\Jtwo{} and
1.2\mm{} continuum observations and data reductions can be found
in~\citet{hilyblant05}. We estimate the absolute position accuracy to be
$3''$.

\FigCuts{} %

We also use a small part of the \HthCOp{} (\Jone{} and \Jthr{}) data, which
were obtained with the \IRAM{} \PdBI{} and 30m telescopes. The whole data
set will be comprehensively described in a forthcoming paper studying the
fractional ionization across the Horsehead edge (Hily-Blant et al. 2007, in
prep). In short, the \HthCOp{}~\Jthr{} line was observed under averaged
winter weather ($\about 3.5\mm$ of water vapor) in rasters along the
direction of the exciting star using the first polarization of the
unrotated HERA.  Each pointing of the rasters was observed in
frequency-switched mode. This resulted in a $140'' \times 75''$ map,
Nyquist sampled along the direction of the exciting star but slightly
undersampled in the orthogonal direction (\ie{} rows separated by $6''$
instead of $4.75''$).  The noise increases quickly at the map edges which
were seen only by a fraction of the HERA pixels.  We finally used a
frequency-switched, on-the-fly map of the \HthCOp{}~\Jone{} line, obtained
at the \IRAMthm{} using the A100 and B100 3mm receivers ($\about 7\mm$ of
water vapor) to produce the short-spacings needed to complement a 7-field
mosaic acquired with the 6 \PdBI{} antennae in the CD configuration
(baseline lengths from 24 to 176~m).

The data processing was done with the \GILDAS{}\footnote{See
  \texttt{http://www.iram.fr/IRAMFR/GILDAS} for more information about the
  \GILDAS{} softwares.} softwares~\citep{pety05b}.  The \IRAMthm{} data
were first calibrated to the \Tas{} scale using the chopper wheel
method~\citep{penzias73}, and finally converted to main beam temperatures
(\Tmb{}) using the forward and main beam efficiencies (\Feff{} \& \Beff{})
displayed in Table~\ref{tab:obs}.  The resulting amplitude accuracy is
\about{} 10\%.  Frequency-switched spectra were folded using the standard
shift-and-add method, after baseline subtraction. The resulting spectra
were finally gridded through convolution by a Gaussian.

\section{Results and discussion}
\label{sec:results}

Figure~\ref{fig:maps} presents the \DCOp~\Jtwo{} and \Jthr{} and the
\CeiO{}~\Jtwo{} integrated intensity maps, together with 1.2\mm{} continuum
emission. All maps are presented in a coordinate system adapted to the
source geometry, as described in the figure caption.  The \DCOp{} emission
is concentrated in a narrow, arc-like structure, delineating the left edge
of the dust continuum emission. A second maximum is found at the extreme
left of the map, associated with a smaller dust continuum peak.
Figure~\ref{fig:cuts} shows the \HthCOp{} and \DCOp{} spectra in a cut
along the direction of the exciting star at $\delta y = 15''$ (horizontal
dashed line of Figure~\ref{fig:maps}). This cut intersects the \DCOp{}
emission peak which is close to the illuminated edge of the nebula ($<50''$
or $\about 0.1\pc$).  To our knowledge, this is the brightest \DCOp{}
emission (4\K{}) detected in an interstellar cloud close to a bright
\HH{}/PAH emitting region~\citep{habart05,pety05a}. The $15''$ spatial
shift between the \DCOp{} and \CeiO{}/continuum emission peaks likely
results from the steep thermal gradient.  The region where the \DCOp{}
emission is produced, is probably cooler than the region where the \CeiO{}
lines and 1.2~mm continuum emission peak, \ie{} cooler than 30\K{}, the
minimum temperature needed to explain the intensity of the \CeiO{}~\Jtwo{}
lines in the cloud edge~\citep{goicoechea06}.

In order to constrain the [\DCOp{}]/[\HCOp{}]
abundance\footnote{$[\DCOp]=n(\DCOp)/n(\HH)$.} ratio from the observed line
emission, we assumed that both species coexist within the same gas
(implying the same physical conditions). This assumption is mainly
justified by the spatial coincidence of the \HthCOp{} and \DCOp{} emission
\emph{peaks} (\ie{} where
$T_\emr{mb}\cbrace{\DCOp{}(2-1)}/T_\emr{mb}\cbrace{\HthCOp{}(1-0)} \simeq
1$).  Besides, we used the \HthCOp{} lines to determine the line--of--sight
\HCOp{} column density.  Indeed, the direct determination of the \HCOp{}
column density from its rotational line emission is hampered by the large
\HCOp{} line opacities and their propensity to suffer from self--absorption
and line scattering effects~\citep{cerni87}.  In addition, large critical
densities for \HCOp{} (and its isotopologues) are expected even for the
lowest--$J$ transitions due to its high dipole moment: $\about 4$~D
(Table~\ref{tab:n_cr}). Hence, thermalization will only occur at very high
densities. For lower densities, $n < n_{crit}$, subthermal excitation
dominates as $J$ increases. Therefore, in order to accurately determine the
mean physical conditions and the [\DCOp{}]/[\HthCOp{}] ratio at the \DCOp{}
peak, we have used a nonlocal, non-LTE radiative transfer code including
line trapping, collisional excitation and radiative excitation by cosmic
background photons~\citep{goicoechea03,goicoechea06}.  Collisional rates of
\HthCOp{} and \DCOp{} with \HH{} and He have been derived from the
\HCOp{}--\HH{} rates of~\citet{flower99}.

Assuming a maximum extinction depth of $\Av\simeq$50 along the
line-of-sight where \DCOp{} peaks~\citep{wardthompson06}, the observed
\DCOp{}~\Jtwo{} and~\Jthr{} line intensities are well reproduced (with line
opacities $\sim$1.5) only if the gas is cold (10--20\K{}) and dense
($n(\HH) \geq 2 \times 10^5\pccm$). This high density is consistent with
the one required to reproduce the CS~\Jfiv{}
excitation~\citep{goicoechea06} and with the value derived from dust submm
continuum emission~\citep{wardthompson06}. The weakness of the
\HthCOp{}~\Jthr{} line compared to the \DCOp{}~\Jthr{} line is caused in
part by its larger Einstein coefficient (a factor \about{}1.7 larger) and
its higher energy level (see Table~\ref{tab:n_cr}).  This implies that the
\HthCOp{}~\Jthr{} line is \textit{more} subthermally excited than the
analogous \DCOp{} line for the derived densities and temperatures.  Note
that we have not included collisions with electrons in this excitation
analysis. In fact, the expected ionization fraction in such a cold and
dense condensation is usually low, $<10^{-7}$ \citep{caselli99}.  The
derived \DCOp{} and \HthCOp{} column densities toward the \DCOp{} peak are
$\simeq (0.5-1) \times 10^{13}\pscm$ (i.e.,
[\DCOp{}]$\simeq$[\HthCOp{}]$\simeq (1-2) \times10^{-10}$).  Assuming a
$^{12}$C/$^{13}$C=60 isotopic ratio \citep{milam05}, we finally find a
$[\DCOp]/[\HCOp] \geq 0.02$ abundance ratio.

\TabTrans{} %


In order to understand the observed deuterium fractionation in the dense
gas close to the Horsehead PDR, we have modeled the steady state deuterium
gas phase chemistry in a cloud with a proton density
$n_\emr{H}=n(\emr{H})+2n(\HH) = 4 \times 10^5\pccm$ illuminated by a FUV
field 60 times the mean interstellar radiation field.  We used the
\texttt{Meudon PDR code}\footnote{Publicly available at
  \texttt{http://aristote.obspm.fr/MIS/}}, a photochemical model of a
unidimensional PDR~\citep[see][for a detailed
description]{lebourlot93,lepetit06} and its associated chemical reaction
network. As this network only includes singly deuterated species, we added
the \DD{} and \HDDp{} species and associated reactions
from~\citet{flower06}.  Nevertheless, these additional reactions do not
affect much the predicted \DCOp{} abundances. Only \HH{}, HD and \DD{} form
on grain surfaces because the used chemical network allows only H and D
atoms to accrete onto dust grains. We chose the following gas phase
abundances: D/H=1.6$\times$10$^{-5}$, He/H=0.1, O/H=3$\times$10$^{-4}$,
C/H=1.4$\times$10$^{-4}$, N/H=8$\times$10$^{-5}$, N/H=8$\times$10$^{-5}$,
S/H=3.5$\times$10$^{-6}$ \citep{goicoechea06}, Si/H=1.7$\times$10$^{-8}$,
Na/H=2.3$\times$10$^{-9}$ and Fe/H=1.7$\times$10$^{-9}$.

\FigChem{} %

We first investigated the role of gas thermodynamics in the \HCOp{}
deuterium fractionation. To do this, we stopped to solve the thermal
balance when the FUV absorption was large enough so that the temperature
reaches a minimum value that we kept constant in the most shielded regions
of the PDR. Figure~\ref{fig:chem} shows the predicted temperature profiles
as well as the [\HHDp{}]/[\HHHp{}] and [\DCOp{}]/[\HCOp{}] abundance ratios
as a function of the cloud depth for a minimum value of T$_k$=15, 20, 30
and 60\K{}.  The predicted [\DCOp{}]/[\HCOp{}] ratio scales with the
[\HHDp{}]/[\HHHp{}] ratio, as expected when \DCOp{} gets fractionated by
the reaction of CO with \HHDp{} and mainly destroyed by dissociative
recombination with electrons~\citep[see \eg{}][]{guelin82}.  The displayed
models also imply that low gas temperatures ($\leq 20\K$) are needed to
reproduce the observed [\DCOp{}]/[\HCOp{}] ratio at $\delta x=$40-45$''$
($\Av \sim 10-20$ depending on the assumed density profile). This is easily
understood because the exchange reaction between \HHHp{} and HD is most
efficient at low temperatures~\citep{gerlich02a}.  Therefore, the observed
$[\DCOp]/[\HCOp] \geq 0.02$ abundance ratio can be reproduced using gas
phase chemistry only if the gas cools down from the photodissociation front
to $\leq 20\K$, in good agreement with the \DCOp{} excitation calculations.
Note, however, that CO freeze--out is believed to further enhance the
[\DCOp{}]/[\HCOp{}] ratio over the values predicted by pure gas phase
fractionation by increasing the abundance of \HHHp{} and
\HHDp{}~\citep[]{brown89a,caselli99}. The \CeiO{}~\Jtwo{} emission shown in
Figure~\ref{fig:cuts} substantially decreases at the \DCOp{} peak.  This
behavior is reminiscent of CO depletion but it could also come from a
combination of lower excitation and of opacity effects.  Future
observations of molecular tracers of gas depletion are needed to constrain
the dominant scenario.

The \DCOp{} lines stay undetected in the warm gas where \HCOp{} (not shown
here) and \HthCOp{} still emit. Indeed, \DCOp{} can not be abundant in the
photodissociation front, where the large photoelectric heating rate implies
warm temperatures ($T_k > 50\K$), because the reaction of \HHDp{} with
\HH{} dominates and implies a low \HHDp{} abundance ($[\HHDp{}]/[\HHHp{}]
\simeq 2 \times 10^{-4}$). From the upper limit of the \DCOp{} emission at
$\delta x=$10-15$'' (\Av \sim1)$, we estimate a low abundance ratio
$[\DCOp]/[\HCOp] < 10^{-3}$ in the FUV photodominated gas, in agreement
with the model predictions.


The small distance to the Horsehead nebula ($\sim 400\pc$), its low FUV
illumination and its high gas density imply that many physical and chemical
processes, with typical gradient lengthscales ranging between $1''$ and
$10''$, can be probed in a small field-of-view (less than $50''$).  The
Horsehead edge thus offers the opportunity to study in great detail the
transition from the warmest gas, dominated by photodissociation processes
and photoelectric heating, to the coldest and shielded gas where strong
deuterium fractionation is taking place.  Therefore, the Horsehead edge is
the kind of source needed to serve as a reference for PDR
models~\citep{pety06} and offers a realistic template to analyze more
complex galactic or extragalactic sources.

\begin{acknowledgements}
  We thank M. Guelin for useful comments and the IRAM PdBI and 30m staff
  for their support during the observations. JRG was supported by an
  \textit{individual Marie Curie fellowship}, contract MEIF-CT-2005-515340.
\end{acknowledgements}

\bibliography{dcop}%
\bibliographystyle{aa}%

\end{document}